\documentclass[5p,number,sort&compress,times,fleqn]{elsarticle}
\usepackage{graphicx}

\newcommand{\half}{{\textstyle \frac{1}{2}}}
\newcommand{\bfsfI}{\mbox{\sffamily\bfseries{I}}}
\newcommand{\bfsfG}{\mbox{\sffamily\bfseries{G}}}
\newcommand{\bfsff}{\mbox{\sffamily\bfseries{f}}}
\newcommand\bfmu{\mbox{\boldmath $\mu$}}

\begin{document}

\title{Quantized Media with Absorptive Scatterers and Modified Atomic Emission Rates}

\author{L.G.~Suttorp and A.J.~van~Wonderen}

\address{Instituut voor Theoretische Fysica, Universiteit van Amsterdam,
Science Park 904, 1098 XH Amsterdam, The Netherlands}

\begin{abstract} 
  Modifications in the spontaneous emission rate of an excited atom that
  are caused by extinction effects in a nearby dielectric medium are
  analyzed in a quantummechanical model, in which the medium consists of
  spherical scatterers with absorptive properties. Use of the dyadic Green
  function of the electromagnetic field near a a dielectric sphere leads to
  an expression for the change in the emission rate as a series of
  multipole contributions for which analytical formulas are obtained. The
  results for the modified emission rate as a function of the distance
  between the excited atom and the dielectric medium show the influence of
  both absorption and scattering processes.
\end{abstract}



\maketitle

\setlength{\mathindent}{0cm}

\renewcommand{\theequation}{\thesection.\arabic{equation}}

\section{Introduction}\label{sec1}

The emission rate of an excited atom is modified if the electromagnetic
properties of its surroundings differ from that of vacuum \cite{P46}.  For
an atom in front of a dielectric medium filling a half-space, the rate
varies with the distance between the atom and the medium \cite{A75} -
\cite{KSW01}. Usually, the medium is taken to be homogeneous on the scale
of the atomic wavelength, so that its electromagnetic properties are fully
described by a susceptibility, which does not vary appreciably on the scale
of the wavelength. In general, it will be complex to account for absorption
and dispersion. For such a configuration modifications of the atomic
radiative properties have been confirmed experimentally a few years ago
\cite{ICLS04}.

If the structure of the medium cannot be neglected, scattering effects play
a role as well, so that extinction in such a medium is driven by both
absorption and scattering. Extinction by scattering in material media is
quite common, owing to the presence of impurities and defects. The
interplay between the two types of extinction in atomic decay rates can be
investigated in a model in which both of these features occur
simultaneously.  In a recent paper \cite{SvW10} we have studied the change
in the decay rate of an atom in the presence of a medium consisting of
non-overlapping spheres that are made of absorptive dielectric
material. The spheres are distributed randomly in a half-space, with a
uniform average density.  In order to describe the absorptive dielectric
material of the spheres in a quantummechanically consistent way a
damped-polariton model has been employed \cite{HB92b}.  By introducing an
effective susceptibility for the composite medium and after a detailed
analysis of surface contributions, we could derive an asymptotic expression
for the change in the emission rate at relatively large distances between
the atom and the medium.

In the present paper we take a somewhat different approach so as to derive
an analytic expression for the emission rate that is valid for all
distances between the atom and the medium. We shall start from exact
expressions for the electromagnetic Green function in the presence of a
dielectric sphere of arbitrary radius. As before, the absorptive dielectric
material of the spheres will be described by a damped-polariton model.  For
simplicity we shall assume that the density of the spherical scatterers in
the medium is low, so that multiple-scattering effects can be neglected.

\section{Spontaneous emission in the presence of absorbing dielectrics}
In the damped-polariton model an absorptive linear dielectric medium is
described by a polarization density that is coupled to a bath of harmonic
oscillators with a continuous range of frequencies~\cite{HB92b}.  The
Hamiltonian of the damped-polariton model can be diagonalized exactly, as
has been shown both for the case of a uniform dielectric \cite{HB92b} and
for a dielectric with arbitrary inhomogeneities
\cite{SWo04}. Diagonalization of the Hamiltonian in the general
non-homogeneous case yields
\begin{equation}
H_d=\int d  {\bf r}\int_0^{\infty} d \omega\, \hbar\omega \, {\bf
C}^{\dagger}({\bf r},\omega)\cdot {\bf C}({\bf r},\omega)\, , \label{2.1}
\end{equation}
with annihilation operators ${\bf C}({\bf r},\omega)$ and associated
creation operators.  The electric field can be expressed in terms of these
operators as \cite{SWo04}:
\begin{equation}
{\bf E}({\bf r})=\int d {\bf r}'\int_0^{\infty} d \omega \,
\bfsff_E({\bf r},{\bf r}',\omega)\cdot{\bf C}({\bf r}',\omega) +{\rm
h.c.} \, ,\label{2.2} 
\end{equation} 
with a tensorial coefficient:
\begin{equation}
\bfsff_E({\bf r},{\bf r}',\omega)=-i\, \frac{\omega^2}{c^2}
\left(\frac{\hbar \, {\rm Im}\,\varepsilon({\bf
r}',\omega+i0)}{\pi\varepsilon_0}\right)^{1/2}\, 
\bfsfG({\bf r},{\bf r}',\omega+i 0) \, . \label{2.3}
\end{equation} 
Here $\varepsilon$ is the complex local (relative) dielectric constant, which follows
from the parameters of the model. Furthermore, $\bfsfG$ is the tensorial
Green function, which satisfies the differential equation
\begin{eqnarray}
 && -\nabla\times [\nabla\times \bfsfG ({\bf r},{\bf
 r}',\omega+i 0)]\nonumber\\
&&+\frac{\omega^2}{c^2}\, \varepsilon({\bf
 r},\omega+i 0)\, 
\bfsfG({\bf r},{\bf r}',\omega+i 0)
=\bfsfI\, \delta({\bf r}-{\bf r}')\, , 
 \label{2.4}
\end{eqnarray}
with $\bfsfI$ the unit tensor. 

The atomic decay rate in the presence of an absorbing dielectric follows
from the inhomogeneous damped-polariton mod\-el in its diagonalized form by
employing perturbation theory in leading order \cite{SvW10}. It can be
expressed as an integral over a product of the coefficients (\ref{2.3}) and
suitable atomic matrix elements:
\begin{eqnarray}
  \Gamma=\frac{2\pi}{\hbar^2 \omega_a^2}
\int d {\bf r} \int d {\bf r}' \int d {\bf r}'' \, 
\langle e|{\bf J}_a({\bf r}')|g\rangle \cdot
\bfsff_E({\bf r}',{\bf r},\omega_a) 
\cdot &&\nonumber\\
\cdot\tilde{\bfsff}_E^\ast({\bf r}'',{\bf r},\omega_a)\cdot 
\langle g|{\bf J}_a({\bf r}'')|e\rangle\,  , &&
\label{2.5}
\end{eqnarray}
with $e$ and $g$ denoting the excited and the ground state of the atom,
$\omega_a$ the atomic transition frequency, and the tilde denoting the
tensor transpose. Furthermore, ${\bf J}_a({\bf r})$ is the atomic local
current density $-\half e\sum_i\{{\bf p}_i/m,\delta({\bf r}-{\bf r}_i)\}$,
with ${\bf r}_i\, , \, {\bf p}_i$ the positions and momenta of the
electrons and curly brackets denoting the anticommutator. In the
electric-dipole approximation the atomic decay rate can be expressed in
terms of the Green function as:
\begin{equation}
\Gamma= -\frac{2\omega_a^2}{\varepsilon_0 \hbar c^2} 
\,\langle e|\bfmu|g\rangle \cdot
{\rm Im}\, \bfsfG({\bf r}_a,{\bf r}_a,\omega_a+i 0)\cdot 
\langle g|\bfmu|e\rangle \, ,
\label{2.6}
\end{equation}
with ${\bf r}_a$ the atomic position and $\bfmu=-e\sum_i({\bf r}_i-{\bf
  r}_a)$ the atomic electric dipole moment. The above expression for the
decay rate of an excited atom in the presence of an inhomogeneous
absorptive dielectric can be obtained as well by invoking the
fluctuation-dissipation theorem \cite{BHLM96,SKW99}.

\section{Green functions}\label{sec2}
\setcounter{equation}{0} The Green function in vacuum fulfils the
differential equation (\ref{2.4}) with $\varepsilon=1$. It follows from the
scalar Green function $G_s({\bf r},{\bf r}',\omega)={\rm exp}(i\omega |{\bf
  r}-{\bf r}'|/c)/(4\pi|{\bf r}-{\bf r}'|)$ as $\bfsfG_0({\bf r},{\bf
  r}',\omega)=-[\bfsfI+(c^2/\omega^2)\nabla\nabla]G_s({\bf r},{\bf
  r}',\omega)$. Its explicit form in spherical coordinates is obtained from
the expansion of $G_s$ in spherical harmonics and spherical Bessel
functions. The ensuing form for the vacuum Green function is \cite{T71,C95}
\begin{eqnarray}
&&\bfsfG_0({\bf r},{\bf r}',\omega+i0)=-ik\sum_{\ell=1}^{\infty}
\sum_{m=-\ell}^{\ell}\frac{(-1)^m}{\ell(\ell+1)}\nonumber\\
&&\times\left\{\theta(r-r')\left[
{\bf M}_{\ell, m}^{(h)}({\bf r}){\bf M}_{\ell,-m}({\bf r'})+
{\bf N}_{\ell, m}^{(h)}({\bf r}){\bf N}_{\ell,-m}({\bf r'}) 
\right]\right.\nonumber\\
&&+\left. \theta(r'-r)\left[
{\bf M}_{\ell, m}({\bf r}){\bf M}_{\ell,-m}^{(h)}({\bf r'})+
{\bf N}_{\ell, m}({\bf r}){\bf N}_{\ell,-m}^{(h)}({\bf r'}) 
\right]\right\}\nonumber\\
&&+k^{-2}\, {\bf e}_r{\bf e}_r\, \delta({\bf r}-{\bf r}')\, ,
\label{3.1}
\end{eqnarray}
with $k=\omega/c$ the wavenumber, ${\bf e}_r$ a unit vector in the
direction of ${\bf r}$ and $\theta(r)$ a step function that equals 1 for
positive and 0 for negative argument.  The vector harmonics are defined as
\begin{eqnarray}
&&{\bf M}_{\ell,m}({\bf r})=\nabla\wedge[{\bf r}\psi_{\ell,m}({\bf r})]\,
,\label{3.2}\\
&&{\bf N}_{\ell,m}({\bf r})=k^{-1}\nabla\wedge[\nabla\wedge[{\bf r}\psi_{\ell,m}({\bf r})]]\,
,\label{3.3}
\end{eqnarray}
where $\psi_{\ell,m}({\bf r})$ stands for $j_\ell(kr)\,
Y_{\ell,m}(\theta,\phi)$, with $j_\ell$ spherical Bessel functions and
$Y_{\ell,m}$ spherical harmonics. The superscripts $(h)$ in (\ref{3.1})
denote the analogous vector harmonics with spherical Hankel functions
$h_\ell^{(1)}$ instead of $j_\ell$.

The expression (\ref{3.1}) may be checked by substitution in the
differential equation (\ref{2.4}).  Differentiation  of the
step functions yields singular terms, which together with the last term
lead to the right-hand side of (\ref{2.4}).  

The Green function in the presence of a dielectric sphere is the sum of the
vacuum Green function and a correction term. For a sphere centered at the
origin the latter has the form
\cite{T71}-\cite{LKLY94}
\begin{eqnarray}
&&\bfsfG_c({\bf r},{\bf r}',\omega+i0)=
k\sum_{\ell=1}^{\infty}\sum_{m=-\ell}^{\ell}
\frac{(-i)^{\ell}(-1)^m}{2\ell+1}
\nonumber\\
&&\times\left[
B^e_\ell\, {\bf N}_{\ell,m}^{(h)}({\bf r}){\bf N}_{\ell,-m}^{(h)}({\bf r}')+
B^m_\ell\, {\bf M}_{\ell,m}^{(h)}({\bf r}){\bf M}_{\ell,-m}^{(h)}({\bf r}')
\right] 
\label{3.4}
\end{eqnarray}
for ${\bf r}$ and ${\bf r}'$ both outside the sphere.  The electric and
magnetic multipole amplitudes read \cite{M08}-\cite{BW99}:
\begin{equation}
B^p_\ell=i^{\ell+1}\, \frac{2\ell+1}{\ell(\ell+1)}\,
\frac{N^p_\ell}{D^p_\ell}\, ,
 \label{3.5}
\end{equation}
with $p=e,m$. The numerators and denominators are given as
\begin{eqnarray}
  &&N^e_\ell=\varepsilon\, f_\ell(q)\, j_\ell(q')- j_\ell(q)\,f_\ell(q')\, 
  , \nonumber\\
&& N^m_\ell=f_\ell(q)\, j_\ell(q')-j_\ell(q)\,f_\ell(q')\, 
  , \nonumber\\ 
&&D^e_\ell=\varepsilon\, f^{(h)}_\ell(q)\, j_\ell(q')-h^{(1)}_\ell(q)\,f_\ell(q')\, 
  , \nonumber\\
&&D^m_\ell= f^{(h)}_\ell(q)\, j_\ell(q')-h^{(1)}_\ell(q)\,f_\ell(q')\,   , 
\label{3.6}
\end{eqnarray}
with $f_\ell(q)=(\ell+1)\, j_\ell(q)-q\, j_{\ell+1}(q)$ and $f^{(h)}_\ell(q)=(\ell+1)\, h^{(1)}_\ell(q)-q\,
h^{(1)}_{\ell+1}(q)$. The spherical Bessel and Hankel functions depend on $q=k a$ and
$q'=\sqrt{\varepsilon}\, q$, with $a$ the radius of the sphere.

To determine the change in the atomic decay rate due to the presence of a
dielectric sphere one needs the components of the Green function $\bfsfG_c$
for coinciding arguments. The non-vanishing components follow from
(\ref{3.4}) as:
\begin{eqnarray}
&&{\bf e}_r\cdot \bfsfG_c({\bf r},{\bf r},\omega+i0)\cdot {\bf e}_r =
\nonumber\\
&&
=\frac{1}{4\pi k r^2}\sum_{\ell=1}^\infty (-i)^\ell\,  [\ell(\ell+1)]^2\,
B^e_\ell \, [h_\ell^{(1)}(kr)]^2
\label{3.7}
\end{eqnarray}
and
\begin{eqnarray}
&&{\bf e}_\theta\cdot \bfsfG_c({\bf r},{\bf r},\omega+i0)\cdot {\bf
  e}_\theta =
{\bf e}_\phi\cdot \bfsfG_c({\bf r},{\bf r},\omega+i0)\cdot {\bf e}_\phi =
\nonumber\\
&&
=\frac{1}{8\pi k r^2}\sum_{\ell=1}^\infty (-i)^\ell\, \ell(\ell+1) \left\{
B^e_\ell \,
\left[\frac{d}{dr}[rh_\ell^{(1)}(kr)]\right]^2\right.\nonumber\\
&&\left. \rule{3.5cm}{0cm}+
 B^m_\ell\, \left[kr\,  h_\ell^{(1)}(kr) \right]^2
\right\}\, ,
\label{3.8}
\end{eqnarray}
in agreement with \cite{DKW01}. Here ${\bf e}_r$, ${\bf e}_\theta$ and
${\bf e}_\phi$ are unit vectors in a spherical coordinate system.

\section{Decay near a half-space of absorptive scatterers}
\setcounter{equation}{0} 

We consider a halfspace $z<0$ filled with a dilute set of spherical
scatterers. The non-overlapping spheres are randomly distributed with a
uniform average density. An excited atom is located at ${\bf
  r}_a=(0,0,z_a)$, with $z_a>a$, so that the minimal distance between the
atom and the scatterers is positive. The decay rate is given by the sum of
the vacuum decay rate and a correction term. The modified rate depends on
the orientation of the dipole-moment transition matrix element. If the
dipole moment is oriented perpendicular to the $z$-axis, the vacuum rate is
$\Gamma_{0,\perp}=\omega_a^3\, |\langle e|\bfmu_\perp|g\rangle|^2/(3\pi
\varepsilon_0 \hbar c^3)$. A similar formula is valid for a dipole moment
oriented parallel to the $z$-axis, with $\bfmu_\perp$ replaced by
$\bfmu_\parallel$.

If multiple-scattering effects are neglected, the correction term in the
decay rate is given by the sum of the correction terms due to all
spheres. For the perpendicular orientation one finds
\begin{eqnarray} 
&&\Gamma_{c,\perp}= -\frac{2\omega_a^2}{\varepsilon_0 \hbar c^2} 
\,\sum_i\langle e|\bfmu_\perp|g\rangle \cdot\nonumber\\
&&\cdot
{\rm Im}\, \bfsfG_c({\bf r}_a-{\bf R}_i,{\bf r}_a-{\bf R}_i,\omega_a+i 0)\cdot 
\langle g|\bfmu_\perp|e\rangle\, ,
\label{4.1}
\end{eqnarray}
with $\bfsfG_c$ given by (\ref{3.4}) and ${\bf R}_i$ the positions of the
centers of the spheres.  Choosing the $x$-axis to be parallel to the
transition matrix element and averaging over the positions of the spheres
we get
\begin{eqnarray}
&&\langle\Gamma_{c,\perp}\rangle=-\frac{6\pi n c}{\omega_a}\,
\Gamma_{0,\perp}\, {\rm Im} \int_{z<0} d{\bf r}
\nonumber\\
&&\times
\, {\bf e}_x\cdot \bfsfG_c({\bf r}_a-{\bf r},{\bf r}_a-{\bf
  r},\omega_a+i0)\cdot{\bf e}_x\, , 
\label{4.2}
\end{eqnarray}
with $n$ the uniform density of the spheres and ${\bf e}_x$ a unit vector 
along the $x$-axis. The volume integral can be written as
a triple integral, viz.\ over $|{\bf r}-{\bf r}_a|$, $z$ and an azimuthal
angle. Upon carrying out the latter two of these integrals one finds for
the volume integral in (\ref{4.2}):
\begin{eqnarray}
&&\pi \int_{z_a}^\infty dr\left[ 
\left(\frac{z_a^3}{3r}-z_a r+\frac{2r^2}{3}\right)
{\bf e}_r\cdot \bfsfG_c({\bf r},{\bf r},\omega_a+i0)\cdot {\bf e}_r \right.\nonumber\\
&&\left. +\left(-\frac{z_a^3}{3r}+\frac{r^2}{3}\right) 
{\bf e}_\theta\cdot \bfsfG_c({\bf r},{\bf r},\omega_a+i0)\cdot {\bf
  e}_\theta \right.\nonumber\\
&&\left.+\left(-z_ar+r^2\right)
{\bf e}_\phi\cdot \bfsfG_c({\bf r},{\bf r},\omega_a+i0)\cdot {\bf
  e}_\phi\right]\, .
\label{4.3}
\end{eqnarray}
 Insertion of (\ref{3.7}) and (\ref{3.8}) yields
\begin{eqnarray}
&&\langle\Gamma_{c,\perp}\rangle=-\frac{3\pi n c^3}{4\omega_a^3}\,
\Gamma_{0,\perp}\, {\rm Im}
\sum_{\ell=1}^{\infty} (-i)^\ell \ell(\ell+1)\nonumber\\
&&\times\left[B^e_{\ell}\,
  J^e_{\ell,\perp}(\zeta_a)+
B^m_{\ell}\, J^m_{\ell,\perp}(\zeta_a)\right]\, ,\label{4.4}
\end{eqnarray}
with multipole amplitudes $B_\ell^p$ given by (\ref{3.5})-(\ref{3.6}) with
$k=\omega_a/c$, and with the integrals
\begin{eqnarray}
&& J^e_{\ell,\perp}(\zeta)=2\ell(\ell+1)\int_{\zeta}^\infty dt\,
\left(\frac{\zeta^3}{3t^3}-\frac{\zeta}{t}+\frac{2}{3}\right) \left[
  h^{(1)}_\ell(t)\right]^2\nonumber\\
&&+\int_{\zeta}^\infty dt\,
\left(-\frac{\zeta^3}{3t^3}-\frac{\zeta}{t}+\frac{4}{3}\right) \left[
  \frac{d}{dt}\left[th^{(1)}_\ell(t)\right]\right]^2
\label{4.5}
\end{eqnarray}
and
\begin{eqnarray}
&&J^m_{\ell,\perp}(\zeta)=\int_{\zeta}^\infty dt\,
\left(-\frac{\zeta^3}{3t}-\zeta t+\frac{4t^2}{3}\right) \left[
  h^{(1)}_\ell(t)\right]^2\, ,
\label{4.6}
\end{eqnarray}
with $\zeta$ equal to $\zeta_a=(\omega_a+i0) z_a/c$.  The derivative of the spherical
Hankel function in (\ref{4.5}) can be rewritten in terms of Hankel
functions with a different index \cite{AS65}.  For large $\zeta$ the
asymptotic forms of these integrals are
\begin{equation}
J^e_{\ell,\perp}(\zeta)\simeq
(-1)^{\ell+1}\frac{e^{2i\zeta}}{2\zeta} \, , \quad 
J^m_{\ell,\perp}(\zeta)\simeq
(-1)^{\ell}\frac{e^{2i\zeta}}{2\zeta} \, , 
\label{4.7}
\end{equation}
so that (\ref{4.4}) becomes
\begin{equation}
\langle\Gamma_{c,\perp}\rangle\simeq \frac{3\pi n c^3}{8\omega_a^3\zeta_a}\,
\Gamma_{0,\perp}\, {\rm Im}
\sum_{\ell=1}^{\infty} i^\ell \ell(\ell+1)
\left(B^e_{\ell} -
B^m_{\ell}\right)\, e^{2i\zeta_a}\, ,
 \label{4.8}
\end{equation}
which falls off proportionally to $1/\zeta_a$. Substituting the leading terms of
$B^e_1$, $B^e_2$ and $B^m_1$ for small values of $q$ and $\varepsilon-1$
one recovers a result found before \cite{SvW10}.

Similar expressions may be obtained for the correction to the decay rate of
an excited atom with a dipole moment parallel to the $z$-axis. Instead of
(\ref{4.2}) one gets a formula with the $zz$-component of $\bfsfG_c$ . Upon
carrying out the integrals one arrives at the analogue of (\ref{4.4}), with
the integrals 
\begin{eqnarray}
&& J^e_{\ell,\parallel}(\zeta)=2\ell(\ell+1)\int_{\zeta}^\infty dt\, 
\left(-\frac{2\zeta^3}{3t^3}+\frac{2}{3}\right) \left[
  h^{(1)}_\ell(t)\right]^2\nonumber\\
&&+\int_{\zeta}^\infty dt\,
\left(\frac{2\zeta^3}{3t^3}-\frac{2\zeta}{t}+\frac{4}{3}\right) \left[
  \frac{d}{dt}\left[th^{(1)}_\ell(t)\right]\right]^2
\label{4.9}
\end{eqnarray}
and
\begin{eqnarray}
&&J^m_{\ell,\parallel}(\zeta)=\int_{\zeta}^\infty dt\,
\left(\frac{2\zeta^3}{3t}-2\zeta t+\frac{4t^2}{3}\right) \left[
  h^{(1)}_\ell(t)\right]^2 \, .
\label{4.10}
\end{eqnarray}
For large $\zeta$ their asymptotic forms are
\begin{equation}
J^e_{\ell,\parallel}(\zeta)\simeq
(-1)^{\ell+1}\frac{ie^{2i\zeta}}{2\zeta^2} \, , \quad
J^m_{\ell,\parallel}(\zeta)\simeq
(-1)^{\ell}\frac{ie^{2i\zeta}}{2\zeta^2}
\, , 
\label{4.11}
\end{equation}
so that the decay rate for large $\zeta_a$ becomes
\begin{eqnarray}
&&\langle\Gamma_{c,\parallel}\rangle\simeq \frac{3\pi n c^3}{8\omega_a^3\zeta_a^2}\,
\Gamma_{0,\parallel}\, {\rm Im}
\sum_{\ell=1}^{\infty} i^{\ell+1} \ell(\ell+1)
\left(B^e_{\ell} -
B^m_{\ell}\right)\, e^{2i\zeta_a}\, .\nonumber\\
&&\mbox{} \label{4.12}
\end{eqnarray}
In contrast to (\ref{4.8}) the right-hand side is proportional to the
inverse square of $\zeta_a$. For general values of $\zeta_a$ we have to
evaluate the integrals in (\ref{4.5})-(\ref{4.6}) and
(\ref{4.9})-(\ref{4.10}), as will be done in the following section.

\section{Evaluation of integrals}
\setcounter{equation}{0}

The integrals $J^p_{\ell,\perp}$ and $J^p_{\ell,\parallel}$ are linear
combinations of integrals of the general form
\begin{eqnarray}
&& I_{\ell_1,\ell_2,n}(\zeta)=\int_1^\infty du\, u^{-n}\, 
h^{(1)}_{\ell_1}(\zeta u)\, h^{(1)}_{\ell_2}(\zeta u)\, ,
\label{5.1}
\end{eqnarray}
which is symmetric in $\ell_1,\ell_2$.  In fact, upon inspecting
(\ref{4.5})-(\ref{4.6}) and (\ref{4.9})-(\ref{4.10}) we find that explicit
expressions are needed for the integrals $I_{\ell,\ell,n}(\zeta)$ with
$n=-2, -1, 0, 1, 3$ and for $I_{\ell,\ell-1,n}(\zeta)$ for $n=-1,0,2$.
With the use of standard identities \cite{AS65} for spherical Hankel
functions and by means of a partial integration we may derive several
relations connecting these integrals for different values of the
parameters:
\begin{eqnarray}
&& I_{\ell_1-1,\ell_2,n}(\zeta)+I_{\ell_1+1,\ell_2,n}(\zeta)=
\frac{2\ell_1+1}{\zeta}\, I_{\ell_1,\ell_2,n+1}(\zeta)\, , \label{5.2}\\
&&(n-\ell_1-\ell_2)\,I_{\ell_1-1,\ell_2,n}(\zeta)+
(n+\ell_1-\ell_2+1)\, I_{\ell_1+1,\ell_2,n}(\zeta)\nonumber\\
&&+(2\ell_1+1)\, I_{\ell_1,\ell_2+1,n}(\zeta)
=\frac{2\ell_1+1}{\zeta}\,h^{(1)}_{\ell_1}(\zeta)\,
h^{(1)}_{\ell_2}(\zeta)\, .
\label{5.3}
\end{eqnarray}  

In order to obtain explicit expressions for $I_{\ell_1,\ell_2,n}$ with
$n=0,1,3$ we start from a result \cite{AS65} that is valid for $n=0$ and
$\ell_1\neq\ell_2$:
\begin{eqnarray}
&&(\ell_1+\ell_2+1)I_{\ell_1,\ell_2,0}(\zeta)=\nonumber\\
&&=\frac{\zeta}{\ell_1-\ell_2}
\left( 
h^{(1)}_{\ell_1}h^{(1)}_{\ell_2-1}-h^{(1)}_{\ell_1-1}h^{(1)}_{\ell_2}\right)
+h^{(1)}_{\ell_1}h^{(1)}_{\ell_2}
\, , \label{5.4}
\end{eqnarray}
as  may be checked by differentiation. We omit the argument
$\zeta$ of the spherical Hankel functions from now on.
To obtain the corresponding expression for $\ell_1=\ell_2$ we put
$\ell_1=\ell+1$, $\ell_2=\ell$ and $n=0$ in (\ref{5.3}) and use (\ref{5.4})
in the second term. In this way we obtain a recursion relation connecting
$I_{\ell,\ell,0}$ for consecutive values of $\ell$. Solving this relation
by employing the identity $I_{0,0,0}(\zeta)=-(2i/\zeta)\,
E_1(-2i\zeta)+[h^{(1)}_0]^2$ (with $E_1$ the exponential integral
\cite{AS65}) as an initial condition,  we find
\begin{eqnarray}
 &&I_{\ell,\ell,0}(\zeta)=-\frac{2i}{(2\ell+1)\zeta}E_1(-2i\zeta)
 \nonumber\\
&&+\frac{2}{2\ell+1}\sum_{k=0}^{\ell}\left[h_k^{(1)}\right]^2-
\frac{1}{2\ell+1}\left[h_\ell^{(1)}\right]^2  \label{5.5}
\end{eqnarray}
for all $\ell\geq 0$. The exponential integral of purely imaginary argument
can be expressed in terms of sine and cosine integrals as
$E_1(-2i\zeta)=-{\rm Ci}(2\zeta)-i{\rm Si}(2\zeta)+i\pi/2$. With the help
of the identities (\ref{5.2}), (\ref{5.4}) and the recursion relations for
the spherical Hankel functions one derives expressions for
$I_{\ell,\ell,1}$ (with $(\ell\geq 1$) and $I_{\ell,\ell,3}$ (with
$\ell\geq 2$), in the form of linear combinations of products of spherical
Hankel functions:
\begin{eqnarray}
&&I_{\ell,\ell,1}(\zeta)=\left[-\frac{\zeta^2}{2\ell(\ell+1)}+\frac{1}{2(\ell+1)}\right]
\left[h_\ell^{(1)}\right]^2\nonumber\\
&&-\frac{\zeta^2}{2\ell(\ell+1)}\left[h_{\ell-1}^{(1)}\right]^2
+\frac{\zeta}{\ell+1}h_\ell^{(1)}h_{\ell-1}^{(1)} 
\label{5.6}
\end{eqnarray}
for $\ell\geq 1$, and 
\begin{eqnarray}
&&I_{\ell,\ell,3}(\zeta)=\left[-\frac{\zeta^4}{3(\ell-1)\ell(\ell+1)(\ell+2)}
-\frac{\zeta^2}{6(\ell+1)(\ell+2)}\right.\nonumber\\
&&\left.+\frac{1}{2(\ell+2)}\right]
\left[h_\ell^{(1)}\right]^2
+\left[-\frac{\zeta^4}{3(\ell-1)\ell(\ell+1)(\ell+2)}\right.\nonumber\\
&&\left.-\frac{\zeta^2}{6(\ell-1)(\ell+2)}\right]\left[h_{\ell-1}^{(1)}\right]^2
+\left[\frac{2\zeta^3}{3(\ell-1)(\ell+1)(\ell+2)}\right.\nonumber\\
&&\left.+\frac{\zeta}{3(\ell+2)}\right]h_\ell^{(1)}h_{\ell-1}^{(1)} 
\label{5.7}
\end{eqnarray}
for $\ell\geq 2$. It turns out that these
formulas cannot be used for $I_{0,0,1}$ and $I_{1,1,3}$, since the
expressions diverge in these cases. However, these special cases can be
obtained straightforwardly by connecting them to the exponential integral
of the same argument as in (\ref{5.5}).

Expressions for $I_{\ell,\ell,n}$ with $n=-2$ follow by choosing
$\ell_1=\ell+1$, $\ell_2=\ell$ and $n=-2$ in (\ref{5.3}). The second term
at the left-hand side drops out for these values of the parameters. As a
result a simple recurrence relation for $I_{\ell,\ell,-2}$ is found, which
may be solved for all $\ell\geq 0$ by employing the identity 
$I_{0,0,-2}(\zeta)=-ie^{2i\zeta}/(2\zeta^3)$ as a starting point. One gets
for $\ell\geq 0$:
\begin{eqnarray}
&&I_{\ell,\ell,-2}(\zeta)=-\frac{1}{2} \left[h_\ell^{(1)}\right]^2-
\frac{1}{2}\left[h_{\ell+1}^{(1)}\right]^2+\frac{2\ell+1}{2\zeta}
h_\ell^{(1)}h_{\ell+1}^{(1)} \, .
\label{5.8}
\end{eqnarray}
Furthermore, by choosing in (\ref{5.3}) the parameters as $n=-2$ and 
$\ell_1=\ell_2$ as either $\ell$ or $\ell+1$, one gets two identities,
which may be combined with (\ref{5.2}) so as to obtain a recursion relation
for $I_{\ell,\ell,-1}$. Solving that relation with the initial condition
$I_{0,0,-1}(\zeta)=-\zeta^{-2}E_1(-2i\zeta)$, we find for all $\ell\geq 0$:
\begin{eqnarray}
&&I_{\ell,\ell,-1}(\zeta)=-\zeta^{-2}E_1(-2i\zeta)+\sum_{k=1}^\ell
\frac{2k+1}{2k(k+1)} \left[h_k^{(1)}\right]^2 \nonumber\\
&&+\frac{1}{2} \left[h_0^{(1)}\right]^2-\frac{1}{2(\ell+1)}
\left[h_{\ell}^{(1)}\right]^2\, .
\label{5.9}
\end{eqnarray}
It should be noted that the sum drops out for $\ell=0$. 

Finally, we need expressions for $I_{\ell,\ell-1,p}$ for $p=-1$ and
$p=2$. Once more we use the identity (\ref{5.3}), now for the choice
$\ell_1=\ell_2=\ell$ and $n=-1$. It yields a recursion relation for
$I_{\ell,\ell-1,-1}$ from which we get for $\ell\geq 1$:
\begin{eqnarray}
&& I_{\ell,\ell-1,-1}(\zeta)=-i\zeta^{-2}E_1(-2i\zeta)+\zeta^{-1}
\sum_{k=0}^{\ell-1} \left[h_k^{(1)}\right]^2\, .
\label{5.10}
\end{eqnarray}
Turning to the case $p=2$, one derives a result for $I_{\ell,\ell-1,2}$ by
a repeated use of (\ref{5.2}) in combination with (\ref{5.4}). Once again
linear combinations of products of two spherical Hankel functions are
found, at least for $\ell\geq 2$:
\begin{eqnarray}
&& I_{\ell,\ell-1,2}(\zeta)=\left[-\frac{\zeta^3}{3(\ell-1)\ell(\ell+1)}-\frac{\zeta}{6(\ell+1)}\right]
\left[h_\ell^{(1)}\right]^2\nonumber\\
&&+\left[-\frac{\zeta^3}{3(\ell-1)\ell(\ell+1)}-\frac{\zeta}{6(\ell-1)}\right]\left[h_{\ell-1}^{(1)}\right]^2
\nonumber\\
&&+\left[\frac{2\zeta^2}{3(\ell-1)(\ell+1)}+\frac{1}{3}\right]h_\ell^{(1)}h_{\ell-1}^{(1)}\, . 
\label{5.11}
\end{eqnarray}
For $\ell=1$ this expression is singular and cannot be used. From a direct
evaluation of the integral for this special case one finds that an
exponential integral shows up, as before.

\section{Results}
\setcounter{equation}{0}

The explicit expressions for the basic integrals (\ref{5.1})  that we derived in the previous
section can be employed now to determine the corrections to the decay
rate. For a perpendicular orientation of the excited atom the correction
(\ref{4.4}) to the decay rate is governed by the integrals
(\ref{4.5})-(\ref{4.6}) for which we get upon substitution of the relevant
contributions:
\begin{eqnarray}
&&J^e_{\ell,\perp}(\zeta)=\zeta E_1(-2i\zeta)\nonumber\\
&&
+\left[-\frac{\zeta^5}{6\ell(\ell+1)}-\frac{\zeta^3(2\ell^2-2\ell-3)}{6\ell(\ell+1)}
+\frac{\zeta\ell}{2(\ell+1)}\right]
\left[h_\ell^{(1)}\right]^2\nonumber\\
&&+\left[-\frac{\zeta^5}{6\ell(\ell+1)}-\frac{\zeta^3(2\ell^2+2\ell-3)}{6\ell(\ell+1)}
\right]\left[h_{\ell-1}^{(1)}\right]^2\nonumber\\
&&+\left[\frac{\zeta^4}{3(\ell+1)}+\frac{\zeta^2(2\ell^2+3\ell-2)}{3(\ell+1)}
\right]h_\ell^{(1)}h_{\ell-1}^{(1)}\nonumber\\
&&-\zeta^3\sum_{k=1}^\ell \frac{2k+1}{2k(k+1)}\left[h_k^{(1)}\right]^2
-\frac{1}{2}\zeta^3\left[h_0^{(1)}\right]^2
\label{6.1}
\end{eqnarray}
and
\begin{eqnarray}
&&J^m_{\ell,\perp}(\zeta)=\zeta E_1(-2i\zeta)\nonumber\\
&&
+\left[\frac{\zeta^5}{6\ell(\ell+1)}-\frac{\zeta^3(2\ell+1)}{3(\ell+1)}\right]
\left[h_\ell^{(1)}\right]^2\nonumber\\
&&+\left[\frac{\zeta^5}{6\ell(\ell+1)}-\frac{2\zeta^3}{3}
\right]\left[h_{\ell-1}^{(1)}\right]^2\nonumber\\
&&+\left[-\frac{\zeta^4}{3(\ell+1)}+\frac{2\zeta^2(2\ell+1)}{3}
\right]h_\ell^{(1)}h_{\ell-1}^{(1)}\nonumber\\
&&-\zeta^3\sum_{k=1}^\ell \frac{2k+1}{2k(k+1)}\left[h_k^{(1)}\right]^2
-\frac{1}{2}\zeta^3\left[h_0^{(1)}\right]^2\, , 
\label{6.2}
\end{eqnarray}
for all $\ell\geq 1$. Likewise, the results for a parallel atomic
orientation read
\begin{eqnarray}
&&J^e_{\ell,\parallel}(\zeta)=2\zeta E_1(-2i\zeta)\nonumber\\
&&
+\left[\frac{\zeta^5}{3\ell(\ell+1)}-\frac{\zeta^3(4\ell^2+2\ell-3)}{3\ell(\ell+1)}
+\frac{\zeta\ell}{\ell+1}\right]
\left[h_\ell^{(1)}\right]^2\nonumber\\
&&+\left[\frac{\zeta^5}{3\ell(\ell+1)}-\frac{\zeta^3(4\ell^2+4\ell-3)}{3\ell(\ell+1)}
\right]\left[h_{\ell-1}^{(1)}\right]^2\nonumber\\
&&+\left[-\frac{2\zeta^4}{3(\ell+1)}+\frac{2\zeta^2(4\ell^2+6\ell-1)}{3(\ell+1)}
\right]h_\ell^{(1)}h_{\ell-1}^{(1)}\nonumber\\
&&-\zeta^3\sum_{k=1}^\ell \frac{2k+1}{k(k+1)}\left[h_k^{(1)}\right]^2
-\zeta^3\left[h_0^{(1)}\right]^2
\label{6.3}
\end{eqnarray}
and
\begin{eqnarray}
&&J^m_{\ell,\parallel}(\zeta)=2\zeta E_1(-2i\zeta)\nonumber\\
&&
+\left[-\frac{\zeta^5}{3\ell(\ell+1)}-\frac{2\zeta^3(\ell-1)}{3(\ell+1)}\right]
\left[h_\ell^{(1)}\right]^2\nonumber\\
&&+\left[-\frac{\zeta^5}{3\ell(\ell+1)}-\frac{2\zeta^3}{3}
\right]\left[h_{\ell-1}^{(1)}\right]^2\nonumber\\
&&+\left[\frac{2\zeta^4}{3(\ell+1)}+\frac{2\zeta^2(2\ell+1)}{3}
\right]h_\ell^{(1)}h_{\ell-1}^{(1)}\nonumber\\
&&-\zeta^3\sum_{k=1}^\ell \frac{2k+1}{k(k+1)}\left[h_k^{(1)}\right]^2
-\zeta^3\left[h_0^{(1)}\right]^2\, , 
\label{6.4}
\end{eqnarray}
again for all $\ell\geq 1$.  As remarked above, the exponential integrals
can be expressed in sine and cosine integrals.

After insertion of the expressions (\ref{6.1})-(\ref{6.2}) and the
multipole amplitudes (\ref{3.5}) into (\ref{4.4}), the average correction
to the decay rate for the perpendicular configuration is found in terms of
well-known special functions depending on $\zeta_a$, $q$ and $\varepsilon$. It may
be plotted as a function of $\zeta_a$, for various choices of $q$ and
$\varepsilon$. To facilitate comparison with our previous results \cite{
  SvW10} we introduce the decay rate correction function
$f_\perp(\zeta_a,q,\varepsilon)=-16 \langle\Gamma_{c,\perp}\rangle
/(3nv_0\Gamma_{0,\perp})$ with $v_0=4\pi a^3/3$ the volume of the spheres.

\begin{figure}[h]
  \begin{center}
   \includegraphics[height=4.5cm]{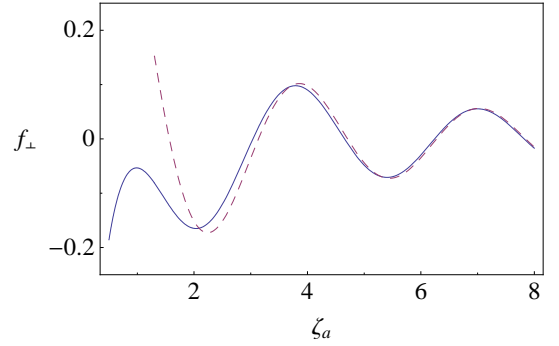}
   \caption{Decay rate correction function $f_\perp(\zeta_a)$ (solid line)
     and its asymptotic form at large distances (dashed line), for a medium
     with scattering spheres (with $q=0.5$, $\varepsilon(\omega_a)=1.5$).}
\label{fig1}
  \end{center}
\end{figure}

\begin{figure}[h]
  \begin{center}
   \includegraphics[height=4.5cm]{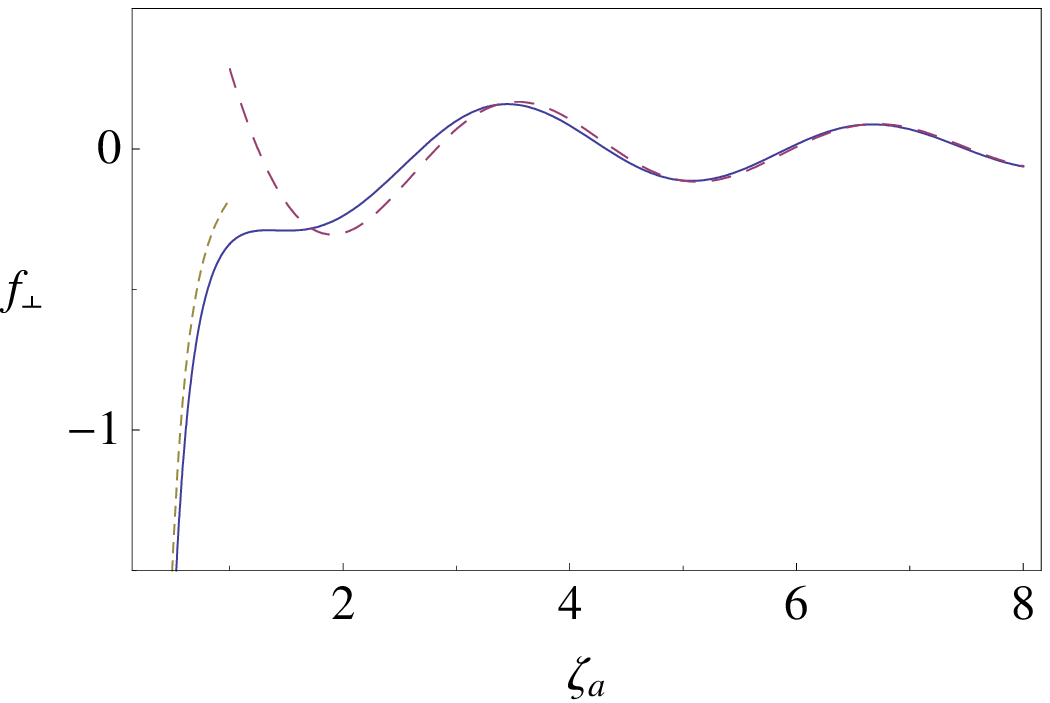}
   \caption{Decay rate correction function $f_\perp(\zeta_a)$ (solid line)
     and its asymptotic forms at small and large distances (dashed lines),
     for a medium with absorbing spheres (with $q=0$,
     $\varepsilon(\omega_a)=1.5+i \, 0.5$).}
\label{fig2}
  \end{center}
\end{figure}

We shall first consider two special cases that we have treated in \cite{SvW10}:
purely scattering spheres and purely absorbing spheres. In the former case
we choose the dielectric constant to be real ($\varepsilon=1.5$) and the
spherical radius to be finite on the scale of the wavelength ($q=0.5$). For
the purely absorbing case with vanishingly small spheres ($q\rightarrow 0$), we take
the dielectric constant to be complex with the value
$\varepsilon=1.5+i\, 0.5$, as in \cite{SvW10}. Since for small $q$ the
multipole amplitudes $B^e_\ell$ and $B^m_\ell$ behave as $q^{2\ell+1}$ and
$q^{2\ell+3}$, respectively, only the electric dipole amplitude
$B^e_1=iq^3(\varepsilon-1)/(\varepsilon+2)$ contributes to (\ref{4.4}) for
the purely absorbing case. In Figs.~\ref{fig1} and \ref{fig2} the curves
for $f_\perp(\zeta_a)$ are compared to their asymptotic counterparts for
large $\zeta_a$ that follow from (\ref{4.8}).  As can be seen from these
figures, the asymptotic curves are quite adequate
already for $\zeta_a\approx 3$.  In the asymptotic regime the results given
in \cite{SvW10} are corroborated. (It should be noted that the curves given
in \cite{SvW10} erroneously represent $-f_\perp$ instead of $f_\perp$.) For
small distances the behaviour of the atomic decay rates in the two cases
differ considerably. In fact, for the purely scattering case of
Fig.~\ref{fig1} the decay rate attains a finite value when $\zeta_a$
approaches its minimum value $q$. On the other hand, for the purely
absorbing case of Fig.~\ref{fig2} the decay rate correction function is governed
by $J^e_{1,\perp}(\zeta)$, which according to (\ref{6.1}) has the asymptotic form
$-1/(4\zeta^3)$ for small $\zeta$. Hence, the decay rate correction
function diverges as $-(3/2)\, {\rm Im}[(\varepsilon-1)/(\varepsilon+2)]
/\zeta_a^3$ for $\zeta_a\rightarrow 0$ in this case.

\begin{figure}[h]
  \begin{center}
   \includegraphics[height=4.5cm]{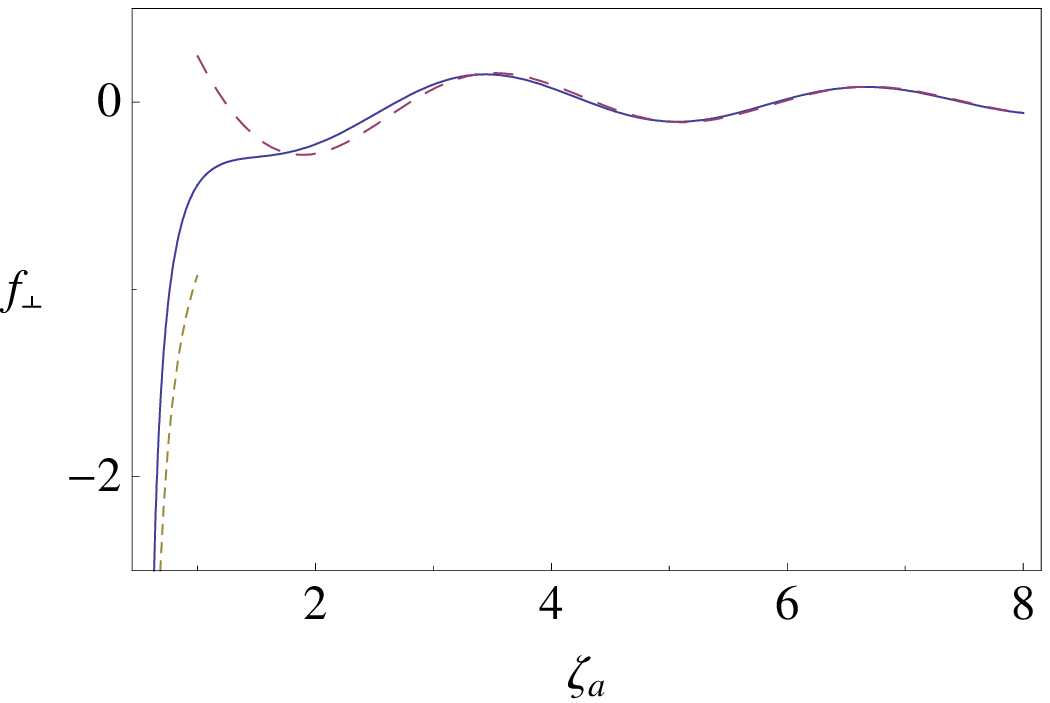}
   \caption{Decay rate correction function $f_\perp(\zeta_a)$ (solid line)
     and its asymptotic forms at small and large distances (dashed lines),
     for a medium with scattering and absorbing spheres (with $q=0.5$,
     $\varepsilon(\omega_a)=1.5+i \, 0.5$).}
\label{fig3}
  \end{center}
\end{figure}

For a more general situation in which both scattering and absorption
take place, we choose $q=0.5$ and $\varepsilon=1.5+i\, 0.5$, with results
presented in Fig.~\ref{fig3}. For large $\zeta_a$ the decay
rate falls off like $\zeta_a^{-1}$, in agreement with (\ref{4.8}). For
small distances the decay rate diverges, as in Fig.~\ref{fig2}.  In fact, as
$\zeta_a \rightarrow q$, the rate is found to be proportional to
$(\zeta_a-q)^{-1}$. This follows from (\ref{4.4}), since the series
converges increasingly slowly when $\zeta_a$ approaches $q$. Indeed, for
large $\ell$ the electric multipole amplitudes $B^e_\ell$ are given by
$[i^\ell/(l^2[(2\ell-1)!!]^2)]\,[(\varepsilon-1)/(\varepsilon+1)]\,
q^{2\ell+1}$, while the integral (\ref{6.1}) gets the form
$-[(2\ell-1)!!]^2/(2\zeta_a^{2\ell+1})$. Hence, the electric multipole
contribution to the $\ell$-th term in the series of (\ref{4.4}) is
$-\half[(\varepsilon-1)/(\varepsilon+1)]\, (q/\zeta_a)^{2\ell+1}$. Since
the magnetic multipole contributions turn out to be negligible for large
$\ell$, the asymptotic form of (\ref{4.4}) for $\zeta_a$ tending to $q$ is
proportional to $\sum_{\ell=1}^\infty (q/\zeta_a)^{2\ell+1}\simeq
q/[2(\zeta_a-q)]$, so that the asymptotic form of $f_\perp$ reads
\begin{equation}
f_\perp\simeq -\frac{3}{4q^2(\zeta_a-q)}\, {\rm Im}
\left[\frac{\varepsilon-1}{\varepsilon+1}\right]
\label{6.5}
\end{equation}
for $\zeta_a\rightarrow q$.  

The physical mechanism for the divergence in the decay rates of
Figs.~\ref{fig2} and \ref{fig3} for small $\zeta_a$ is the efficient
non-radiative energy transfer from the atom to the absorbing spheres that
dominates the atomic decay in the near zone. A similar divergent behaviour
has been found in a classical treatment of the energy transfer between an
excited molecule and a homogeneous absorbing medium filling a halfspace
\cite{CPS74}. It should be noted that (\ref{6.5}) loses its meaning when
$\zeta_a-q$ becomes so small that the approximations made in deriving it
are no longer valid. In particular, perturbation theory in lowest order and
the electric-dipole approximation are not adequate to describe the decay
for very small distances. Furthermore, the notion of scatterers with a
structureless surface gets lost as well in that case.

\begin{figure}[h]
  \begin{center}
   \includegraphics[height=4.5cm]{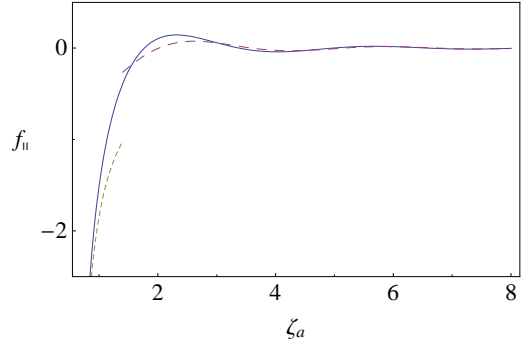}
   \caption{Decay rate correction function $f_\parallel(\zeta_a)$ (solid
     line) and its asymptotic forms at small and large distances (dashed
     lines), for a medium with scattering and absorbing spheres (with
     $q=0.5$, $\varepsilon(\omega_a)=1.5+i\, 0.5$).}
\label{fig4}
  \end{center}
\end{figure}

For the parallel configuration we have likewise evaluated
$f_\parallel(\zeta_a,q,\varepsilon)=-16 \langle\Gamma_{c,\parallel}\rangle
/(3nv_0\Gamma_{0,\parallel})$. The result for the mixed case of both
scattering and absorption is given in Fig.~\ref{fig4} for
the same choice of the parameters $q$ and $\varepsilon$ as in Fig.~\ref{fig3}.
The figure clearly shows that for the parallel configuration the correction
to the atomic decay rate goes faster to zero with increasing $\zeta_a$ than
for the perpendicular configuration. This is in accordance with the
findings of section 4, where it has been seen that in the asymptotic regime
the correction to the atomic decay rate is proportional to the inverse
distance in the perpendicular configuration, but to its square in the
parallel configuration. As before, the asymptotic expression is adequate
from $\zeta_a\approx 3$ onwards. For small distances, with
$\zeta_a\rightarrow q$, the asymptotic form of $f_\parallel$ is twice that
of $f_\perp$, as follows by comparing the asymptotic forms of (\ref{6.1})
and (\ref{6.3}) for large $\ell$.

In conclusion, we have shown how absorption and scattering processes in a
medium may cooperate in modifying the emission rate of an excited atom in
its vicinity. The explicit expressions for the decay rate that we have
obtained permit a detailed analysis of the behaviour of the emission rate
for arbitrary distances between the atom and the medium. As we have seen,
the effects of absorption and of scattering are qualitatively different, when
the atom approaches the medium.

\end{document}